\documentclass[amsmath,amssymb,11pt,superscriptaddress,reprint, preprintnumbers, notitlepage,aps,prl,twocolumn]{revtex4-1}

\pdfoutput=1 
\usepackage[utf8]{inputenc}
\usepackage[english]{babel}
\usepackage{amsmath}
\usepackage{graphicx}
\usepackage{dcolumn}
\usepackage{pbox}
\usepackage{amssymb}
\usepackage{epsfig}
\usepackage[dvipsnames]{xcolor}
\usepackage{slashed}
\usepackage{amssymb}
\usepackage{ mathrsfs }
\usepackage{color}
\usepackage[font=small]{caption}
\usepackage[font=small]{subcaption}
\usepackage{url}
\definecolor{MyDarkBlue}{rgb}{0.1, 0.1, 0.8} 
\definecolor{SBlue}{rgb}{0.2, 0.4, 0.7} 
\definecolor{MyLightBlue}{rgb}{0.22,0.51,0.9}
\definecolor{MyGreen}{rgb}{0.0, 0.5, 0.0}
\definecolor{BrickRed}{rgb}{0.8, 0.25, 0.33}
\RequirePackage{hyperref}
\hypersetup{colorlinks, citecolor=SBlue,linkcolor=MyDarkBlue, urlcolor=PineGreen}

\makeatletter

\makeatletter
\renewcommand\@makecaption[2]{%
  \par
  \vskip\abovecaptionskip
  \begingroup
  
   \small\rmfamily
    \begingroup
     \samepage
     \flushing
     \let\footnote\@footnotemark@gobble
     \@make@capt@title{#1}{#2}\par
    \endgroup
  \endgroup
  \vskip\belowcaptionskip
}
\makeatother

\DeclareUnicodeCharacter{2212}{-}
\begin{document}

\title{\vspace{1cm}\Large Correlating $W$-Boson Mass Shift with Muon ${g-2}$ in the 2HDM}

\author{\bf K.S. Babu}
\email[E-mail: ]{babu@okstate.edu}
\affiliation{Department of Physics, Oklahoma State University, Stillwater, OK, 74078, USA}    

\author{\bf Sudip Jana}
\email[E-mail:]{sudip.jana@mpi-hd.mpg.de}
\affiliation{Max-Planck-Institut f{\"u}r Kernphysik, Saupfercheckweg 1, 69117 Heidelberg, Germany}


\author{\bf Vishnu P.K.}
\email[E-mail:]{ vipadma@okstate.edu}
\affiliation{Department of Physics, Oklahoma State University, Stillwater, OK, 74078, USA}

\begin{abstract}
We show an interesting correlation between the recent high precision measurement of the $W$-boson mass by the CDF collaboration and the muon $(g-2)$ anomaly in the context of the two Higgs doublet model.  One-loop diagrams involving the exchange of neutral scalar bosons can explain the muon $(g-2)$, which however requires significant mass splittings among members of the second Higgs doublet.  These splittings also generate a positive shift in the mass of the $W$-boson, consistent with the recent CDF measurement.  The charged and neutral scalars of the model cannot be heavier than about 600 GeV for a simultaneous explanation of the two anomalies. The entire parameter space of the model can be tested at the LHC by a combination of  same sign dimuon signals in $pp \rightarrow (\mu^+ \mu^+ jj + {E\!\!\!\!/}_{T})$  and  $pp \rightarrow (\mu^+\mu^-\tau^+\tau^-+X)$ signals.

\noindent 
\end{abstract}

\maketitle

\textbf{\emph{Introduction}.--}
A high precision measurement of the $W$ boson mass has recently been reported by the Fermilab CDF collaboration based on high yields of $W$ bosons collected from 2002 to 2011 \cite{CDF:2022hxs}.  The reported value, $M_W = (80,433.5 \pm 9.4)~{\rm MeV}$, differs from the Standard Model prediction of $M_W^{\rm SM} = (80,357 \pm 6)~{\rm MeV}$ \cite{Awramik:2003rn} at the 7 sigma level.  Although the CDF measurement is not fully consistent with LHC measurements (ATLAS measures $M_W({\rm ATLAS}) = (80,370 \pm 19)~{\rm MeV}$ \cite{ATLAS:2017rzl}, LHCb measures $M_W({\rm LHCb}) = 80,354 \pm 32~{\rm MeV}$ \cite{LHCb:2021bjt}) its high precision taken together with the consistent cross checks leading to the $Z$ boson mass in agreement with world average value may be viewed as a tantalizing evidence for new physics beyond the Standard Model (SM).  It is sufficient for such new physics to show up as subtle quantum corrections in order to explain this discrepancy.

Independently, the Fermilab muon ($g-2$) collaboration has reported a precise measurement of the anomalous magnetic moment of the muon to be $a_\mu({\rm FNAL}) = 116592040(54) \times 10^{-11}$ \cite{Muong-2:2021ojo}, in agreement with the prior Brookhaven $(g-2)$ collaboration measurement of $a_\mu({\rm BNL}) = 116592089(63) \times 10^{-11}$ \cite{Muong-2:2006rrc}.  These measurements differ from the standard model theoretical prediction of $a_\mu({\rm theory}) = 116591810(43) \times 10^{-11}$ \cite{Aoyama:2020ynm} at a combined 4.2 sigma confidence level. This anomaly may well be viewed as evidence for new physics.  

It would be intriguing to seek frugal extensions of the SM which simultaneously explain the  upward shift in the $W$ boson mass and the anomalous magnetic moment of the muon.  The purpose of this paper is to show an interesting correlation between the two anomalies in the context of two Higgs doublet model (2HDM).  A simple way to explain the muon $(g-2)$ in the 2HDM is through one-loop diagrams involving neutral Higgs bosons from the second Higgs doublet and the $\tau$-lepton.  Such diagrams have a helicity enhancement proportional to $m_\tau/m_\mu$ in the muon $(g-2)$ and thus can readily explain the BNL and FNAL measurements.  However, these diagrams require significant splittings among the neutral scalar bosons -- in the limit of degenerate masses new contributions to $a_\mu$ would vanish.  We show here that interestingly these mass splittings also provide an upward shift in the mass of the $W$ boson compared to the SM prediction, in agreement with the recent CDF measurement. In presence of such splittings, the process $pp \rightarrow \mu^+ \mu^+ jj + {E\!\!\!\!/}_{T}$ with same-sign dimuons occurs with significant strength and can be used as a test of the model.  Additionally, the process $pp \rightarrow \mu^+ \mu^- \tau^+ \tau^-+X$ is also within reach, given that the scalar particles cannot be heavier than about 600 GeV  in order for the anomalies to be explained. A combination of these two processes can explore the entire parameter space of the model.

\textbf{\emph{Model}.--}  We work within the general 2HDM with no special symmetry imposed (for a review and detailed list of references see Ref. \cite{Branco:2011iw}). In this scenario one can make a field rotation so that only one among the two neutral Higgs boson acquires a vacuum expectation value (VEV). In this Higgs basis, the two Higgs doublets can be parameterized as:
\begin{align} 
	& H_1=\begin{pmatrix}
		G^{+}  \\
		\frac{1}{\sqrt{2}}(v+\phi_1^0+iG^0)    
	\end{pmatrix}   
\quad H_2=\begin{pmatrix}
		H^{+}  \\
		\frac{1}{\sqrt{2}}(\phi_2^0+iA)     
	\end{pmatrix} 
\label{para}
\end{align}
Here $G^+$ and $G^0$ are unphysical Goldstone modes, whereas $H^+$ and $\{\phi_1^0,\phi_2^0, A \}$ belong to the physical spectrum. The VEV of $H_1$, $v\simeq 246$ GeV, governs the electroweak symmetry breaking dynamics. 

The physical scalar masses can be readily computed  from the Higgs potential (see Eq.~(A1) in Appendix). For simplicity we take all the parameters of the potential to be real.  We are also interested in working close to the alignment limit, where the SM-like Higgs $\phi_1^0\approx h$ decouples from the new  CP-even Higgs ($\phi_2^0\approx H$). The properties of $h$ in this limit will not be modified much compared to the SM. In the alignment limit the masses of the physical scalars can be expressed as (see for e.g. Ref. \cite{Babu:2018uik}):
\begin{align}
&m^2_h= \lambda_1v^2,\;
m^2_H=m^2_{22}+\frac{v^2}{2}(\lambda_3+\lambda_4+\lambda_5),
\\
&m^2_A=m^2_H-v^2 \lambda_5,\;
m^2_{H^\pm}=m^2_H-\frac{v^2}{2}(\lambda_4+\lambda_5).
\end{align}
Note that the splittings among the masses of the second doublet are controlled by the quartic couplings $\lambda_4$ and $\lambda_5.$ These splittings will be relevant for both the muon $(g-2)$ anomaly and the $W$ boson mass prediction.

The leptonic Yukawa Lagrangian of the model is given by
\begin{align}
	-&\mathcal{L}_{\text Yuk}\supset  Y \overline{\ell}_L H_1 \ell_R + \widetilde{Y} \overline{\ell}_L H_2 \ell_R + h.c. 
	\label{Yuk1}
\end{align}
The charged lepton mass matrix follows from Eq. (\ref{Yuk1}) as $M_\ell = Y v/\sqrt{2}$.  Without loss of generality we choose the matrix $Y$ to be diagonal, so that Eq. (\ref{Yuk1}) is already in the charged lepton mass eigenbasis.
Analogous couplings exist in the quark sector. While the quark couplings to the $H_1$ field are necessary to generate quark masses, their couplings to $H_2$ field may be taken to be small, or even zero.  In this limit, the second doublet behaves as a leptophilic Higgs, for which low energy constraints from meson decays etc, as well as high energy collider constraints are not that stringent. We shall work in this leptophilic limit of the model.

In addition, in the spirit of explaining the muon $(g-2)$ anomaly, we shall assume the following flavor structure for the Yukawa matrix $\tilde{Y}$ in Eq. (\ref{Yuk1}): 
\begin{equation}
    \tilde{Y}_{\mu\tau} \neq 0,~~\tilde{Y}_{\tau\mu} \neq 0,~\tilde{Y}_{ij} = 0 ~{\rm for~ all~ other}~ i,j~.
    \label{ansatz}
\end{equation}
With this choice, lepton flavor violation constraints are automatically satisfied.  In fact, with this choice a $Z_4$ subgroup of $U(1)_{L_\mu-L_\tau}$ remains unbroken in the Yukawa sector. This $Z_4$ symmetry, if imposed, would set $\lambda_6 = \lambda_7=0$ in the Higgs potential of Eq. (A1). Importantly, this $Z_4$ would admit the $\lambda_5$ coupling, which is crucial for the one-loop diagram for muon $(g-2)$. It is worth noting that the diagonal couplings of the matrix $Y$ of Eq. (\ref{Yuk1}) are allowed by the $Z_4$, so that charged lepton masses are consistently generated. There would be no tree-level flavor-changing neutral currents mediated by neutral Higgs bosons in the limit of exact $Z_4$ symmetry. We shall refer to this scenario as $\mu-\tau$ specific 2HDM.  While this is an interesting limit of the general 2HDM, we shall not explicitly need to adopt this limit in our numerical study.  Rather, the ansatz of Eq. (\ref{ansatz}) would suffice for this purpose.

Muon $(g-2)$ has been studied within the 2HDM by various authors. One class of study uses lepton flavor specific 2HDM where there are no FCNC at tree level \cite{Abe:2015oca,Jana:2020pxx, Wang:2018hnw}. Another possibility that has been studied is muon-specific 2HDM \cite{Abe:2017jqo}. In contrast, the $\mu-\tau$-specific 2HDM that we have explored is closer to the 2HDM studied in Ref. \cite{Omura:2015xcg}.


\textbf{\emph{Muon anomalous magnetic moment}.--}
 \begin{figure}[htb!]
\includegraphics[width=0.35\textwidth]{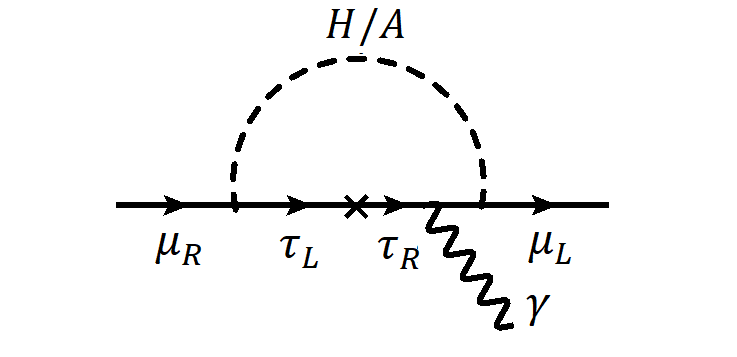}
\caption{The dominant contribution to the muon anomalous magnetic moment.
} 
\label{feynG2}
\end{figure}
In our setup, both neutral scalars and the charged scalar would contribute to the  muon anomalous magnetic moment. 
However, since chirality flip occurs
on the external legs for the charged scalar mediated contribution to  $\Delta a_{\mu}$, the corresponding contribution  is negligible compared to the corrections from the neutral scalars.
The dominant contribution to the muon anomalous magnetic moment in our setup shown in Fig.~\ref{feynG2} which generates a
$\Delta a_{\mu}$ given by \cite{Leveille:1977rc}: 
\begin{align}
\Delta a_{\mu}&=
\frac{ \widetilde{Y}_{\mu\tau} \widetilde{Y}_{\tau\mu}}{16\pi^2}  \left(\frac{m_{\tau}}{m_{\mu}}\right) 
 \left[ G\left(\frac{m_{\tau}^2}{m^2_{\mu}},\frac{m_{H}^2}{m^2_{\mu}} \right)- G\left(\frac{m_{\tau}^2}{m^2_{\mu}},\frac{m_{A}^2}{m^2_{\mu}} \right) \right] 
\label{amuplus}
\end{align}
where the loop function is defined as 
\begin{align} 
&G\left(\omega^a,\omega^b\right)=
\int_0^1 dx \frac{x^2}{ x^2+x(\omega^a-1) +\omega^b(1-x)}.
\label{loopfun}
\end{align} 
Here we have not shown the neutral scalar contributions with external mass flip of the muon, which are much smaller than the contribution shown in Eq. (\ref{amuplus}).  We shall use these expressions for computing $(g-2)$ of the muon. Note that these contributions would require a mass splitting between the neutral scalars $H$ and $A$, since in the degenrate limit their contributions exactly cancel.  

\textbf{\emph{Shift in the \boldmath$W$-boson mass}.--} The $W$-boson mass is calculable within the SM in terms of the precisely measured input parameters $\{G_F, \,\alpha_{\rm em},\, M_Z\}$ as well as the masses of the top quark and the Higgs boson, and the QCD coupling $\alpha_s(M_Z)$ which appear in quantum corrections to $M_W$. The $W$-boson mass is related to these parameters as
\begin{equation}
    M_W^2 \left(1-\frac{M_W^2}{ M_Z^2}\right) = \frac{\pi \alpha_{\rm em}}{\sqrt{2} G_F}(1+\Delta r)
    \label{eq8}
\end{equation}
where $\Delta r$ represents the quantum corrections.  Here $M_W$ and $M_Z$ are the renormalized masses in the on-shell scheme.  From Eq. (\ref{eq8}) $M_W$ can be calculated as
\begin{equation}
    M_W^2 = \frac{M_Z^2}{2}\left[ 1 + \sqrt{1- \frac{4 \pi \alpha_{\rm em}}{ \sqrt{2} G_F M_Z^2}(1+\Delta r)} \right]~.
    \label{MW}
\end{equation}
The parameter $\Delta r$ itself is a function of $M_W^2$, which requires an iterative solution to Eq. (\ref{MW}) to obtain the SM prediction for $M_W$. It is conventional to write $\Delta r$ as
\begin{equation}
    \Delta r = \Delta \alpha -\frac{c_W^2}{s_W^2} \delta \rho + \Delta r_{\rm rem}
\end{equation}
where $\Delta \alpha \simeq 0.06$ arises from the renormalization of $\alpha_{\rm em}$ which is dominated by light fermions, $\delta \rho= T/\alpha(M_Z)$ is the oblique correction with a value of $\Delta r^{[\Delta \rho]} \equiv - (c_W^2/s_W^2) \delta \rho \simeq -0.03$ in the SM primarily arising from top-bottom quark loop diagram, and  $\Delta r_{\rm rem} \simeq 0.01$ is the remaining box and vertex diagram contributions \cite{Lopez-Val:2014jva}.  With the central values of the input parameter, $M_Z = 91.1876$ GeV, $\alpha_{\rm em}^{-1} = 137.036$, $s_W^2 = 0.2315$, $M_t = 172.76$ GeV, $M_h = 125. 25$ GeV, and $G_F = 1.1663787 \times 10^{-5}$ GeV$^{-2}$, one finds $\Delta r^{\rm SM} = 0.038$, and consequently from Eq. (\ref{MW}) $M_W= 80.3564$ GeV. This is 7 $\sigma$ below the CDF value reported recently \cite{CDF:2022hxs}.  This discrepancy may be resolved via quantum corrections that modify $\Delta r$ from its SM value.  Indeed, a value of $\Delta r = 0.03386$ would lead to $M_W = 80. 433$ GeV, the central value from the CDF measurement.  

Assuming that the new physics that modifies $\Delta r$ arises as oblique corrections dominated by $T$-parameter, one would require a new positive contribution to $\delta \rho^{\rm new} = 0.001246$. Using $\alpha(M_Z) = 1/127.935$, this would imply $T = 0.15949$.  The one sigma allowed range for the $T$-parameter consistent with the CDF $W$ boson mass measurement is 
\begin{equation}
    T : \{ 0.159,~ 0.210\}
\end{equation}
which corresponds to the full contribution to $\Delta r$ in the range $\{0.0326,~0.0339\}$.  There is a sub-leading contribution from the oblique $S$-parameter (discussed below), which we also include in our numerical study.  

There are new contributions to the $T$-parameter in the 2HDM, given, in the alignment limit, by
 \begin{align} 
    & T  =  \scriptstyle \dfrac{1}{16\pi^2 \alpha_{\rm em}(M_Z) v^2 } \left\lbrace  { \mathcal{F}(m_{H^+}^2,m_{H}^2) + \mathcal{F}(m_{H^+}^2,m_{A}^2) 
    -\mathcal{F}(m_{H}^2,m_{A}^2) } \right\rbrace \,, \label{eq:T}
\end{align}
where the symmetric function $\mathcal{F}$ is given by
\begin{equation} \label{Fdef}
    \mathcal{F}(m_1^2,m_2^2) \  \equiv \  \frac{1}{2}(m_1^2+m_2^2) -\frac{m_1^2m_2^2}{m_1^2-m_2^2}\ln\left(\frac{m_1^2}{m_2^2}\right)\,.
\end{equation}

In addition to the $T$-parameter contribution, $S$-parameter can also modify the $W$-boson mass. The corresponding shift in the $W$-boson mass is given by \cite{Dugan:1991ck,Peskin:1991sw}
 \begin{align} 
    & \Delta M_W  \simeq - \dfrac{M_{W}\alpha(M_Z)S}{4(c_W^2-s_W^2)}. \label{eq:MW-S}
\end{align}
Here $M_{W}$ is the SM prediction for $M_W$.
In the alignment limit, the $S$-parameter in the 2HDM is given in Ref.~\cite{Barbieri:2006dq} (see Eq.~(B1) in Appendix).
We find that the shift in the $W$- boson mass due to the $S$-parameter is typically small, in the (5-10) MeV range, which we have included in our analysis. 

In our setup $G_F$ appearing in muon decay is modified due to new diagrams affecting the $W$-$\mu$-$\nu_{\mu}$ coupling proportional to the Yukawa coupling $\tilde{Y}_{\mu \tau}$ of Eq. (\ref{ansatz}). This shift in  $G_F$ can be obtained from analogous expressions given in Ref.~\cite{Abe:2017jqo} (see Eq.~(C1) of Appendix). These correction would contribute to $\Delta r$ and thus would modify he $W$-boson mass. However, we find that the shift in the $W$-boson mass arising from these corrections is at most $1$ MeV in our setup, even with the Yukawa coupling $|\tilde{Y}_{\mu\tau}|\simeq 1$. 

\textbf{\emph{Other constraints}.--}
The charged Higgs boson in the model would mediate a new decay for $\tau$ lepton: $\tau^- \rightarrow \mu^- \nu_\mu \overline{\nu}_\tau$. The effective Lagrangian for this decay can be written down (after a Fierz reordering) as 
\begin{equation}
    {\cal L}_{\rm eff} = -\frac{\tilde{Y}_{\mu\tau} \tilde{Y}^*_{\tau\mu}}{2M_{H^+}^2} (\overline{\nu}_{\mu_L} \gamma_\mu \nu_{\tau_L}) (\overline{\mu}_R \gamma^\mu \tau_R) + h.c.
\end{equation}
Note that the $\tau^- \rightarrow \mu^- \nu_\mu \overline{\nu}_\tau$ decay does not interfere with the standard decay $\tau^- \rightarrow \mu^- \overline{\nu}_\mu \nu_\tau$.  Even so, the strength for the decay $\tau \rightarrow \mu \nu \nu$ will differ from that of the usual muon decay which does not receive any new contribution.  From the limit $g_\tau/g_\mu = 1.0011 \pm 0.0015$ obtained from these two decays \cite{Pich:2013lsa}, one obtains 
\begin{equation}
    |\epsilon_\tau| \equiv \left| \frac{\tilde{Y}_{\mu \tau}\tilde{Y}_{\tau \mu}^*}{g^2} \frac{M_W^2}{M_{H^+}^2} \right| \leq 0.089
\end{equation}
at the 2 sigma level. We shall impose this constraint in our numerical analysis, which would serve as the most stringent constraint on the mass of $H^\pm$.  There is also a constraint $|\epsilon_\tau|< 0.213$ from $\tau$ lifetime measurements, which is however weaker.

There are direct search limits on the mass of $H^\pm$ from LEP and LHC experiments.  These limits depend on the branching ratios of $H^+$ into specific lepton flavor. In our scenario $H^+$ would decay into $\tau^+ \nu_\mu$ or $\mu^+ \nu_\tau$.  If the branching ratios for the two decay modes are comparable, $H^\pm$ mass can be as low as about 100 GeV \cite{Babu:2019mfe}, while for a 100\% branching ratio into $\mu^+ \nu$ LHC limits of order 250 GeV become applicable.

The neutral CP-even and CP-odd scalar fields can have mass as low as 100 GeV.  Once produced, these particles would decay dominantly to $\mu^+ \tau^-$ or $\mu^- \tau^+$.  There are no dedicated searches by ATLAS or CMS for resonances decaying this way.  In our analysis we shall allow these masses to be as low as 100 GeV.

Loop corrections to the $W$-$\mu$-$\nu_{\mu}$ vertex proportional to $\tilde{Y}_{\mu\tau}$, as well as the oblique parameters $T$ and $S$ would affect the partial decay width $\Gamma(W\rightarrow \mu \overline{\nu}_\mu )$. These shifts can be obtained from expressions given in  Ref.~\cite{Abe:2017jqo} with appropriate changes in the couplings. For all the benchmark points listed in Table \ref{benchmarkpoints}, we find that the shift in the $\Gamma(W\rightarrow \mu \overline{\nu}_\mu )$ is less than $0.5$ MeV, for the Yukawa coupling $|\tilde{Y}_{\mu\tau}|\simeq 1$. Similarly, the vertex corrections to the $Z$-$\mu$-$\mu$ coupling and the oblique corrections  would modify
the partial decay width $\Gamma(Z \rightarrow\mu \bar{\mu})$. However, such effects are always very small in our setup. For $|\tilde{Y}_{\mu\tau}|\simeq 1$ (or $|\tilde{Y}_{\tau\mu}|\simeq 1$), we find that the shift in the $\Gamma(Z \rightarrow \mu \bar{\mu} )$ to be less than 1 MeV.

\begin{figure}[htb!]
\includegraphics[width=0.45\textwidth]{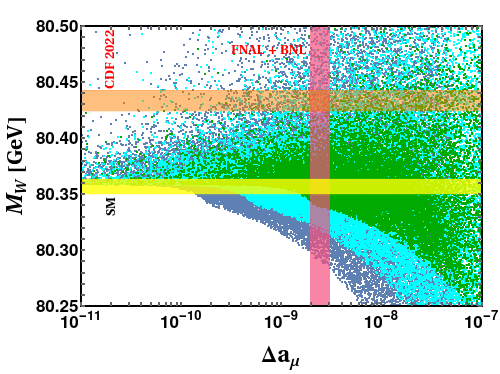}
\caption{ Theoretical predictions and experimental measurements of $W$-boson mass and muon anomalous magnetic moment in the 2HDM. The different colored scattered points depict the correlated predictions for the $W$-boson mass ($M_W$) and muon anomalous magnetic moment ($\Delta a_{\mu}$) in our framework for different choices of Yukawa coupling: $\sqrt{| \tilde{Y}_{\mu\tau}  \tilde{Y}_{\tau\mu} |}=1.0$ (green), 0.5 (cyan), and 0.3 (blue).
} \label{main2}
\end{figure}

{\textbf {\textit {Results.--}}} 
In Fig.~\ref{main2} we have shown a direct correlation between $W$- boson mass and muon anomalous magnetic moment within our framework. As noted earlier, the muon anomalous magnetic moment and the shift in the $W$- boson mass depend on the parameters $(m_H,\lambda_4,\lambda_5,|\tilde{Y}_{\mu\tau}  \tilde{Y}_{\tau\mu} |)$. In Fig.~\ref{main2}, we choose three different values for the $\sqrt{|\tilde{Y}_{\mu\tau}  \tilde{Y}_{\tau\mu} |}$ (1.0, 0.5, and 0.3). We vary the quartic couplings $\lambda_4,\lambda_5$ uniformly in the range $(-\sqrt{4\pi},\sqrt{4\pi})$ and $m_H  \in (110$ GeV - 1 TeV). Then from this data set we choose data points which satisfy $\tau$-decay universality constraints, positivity criteria of the squared masses of the scalar states, and the conditions $m_A, m_{H^{\pm}}\geq 110$ GeV. Our results are shown in Fig. \ref{main2}, where we have plotted $M_W$ versus $\Delta a_\mu$.  Here we also show the FNAL + BNL one sigma range for $\Delta a_\mu$, as well as the 1 sigma range for $M_W$ from the CDF experiment.  The yellow band in Fig. \ref{main2} is the SM prediction for $M_W$.  We see from Fig. \ref{main2} that there are large number of points which explain both anomalies within 1 sigma.

In Table 1 we have listed several  benchmark points for the various scalar masses that satisfy both anomalies, corresponding to three sets of Yukawa couplings.  These benchmark points, as well as the results from our full scan, show that the masses of the scalars cannot exceed about 600 GeV, making the model fully testable at the LHC.

\begin{figure}[htb!]
$$
\includegraphics[width=0.45\textwidth]{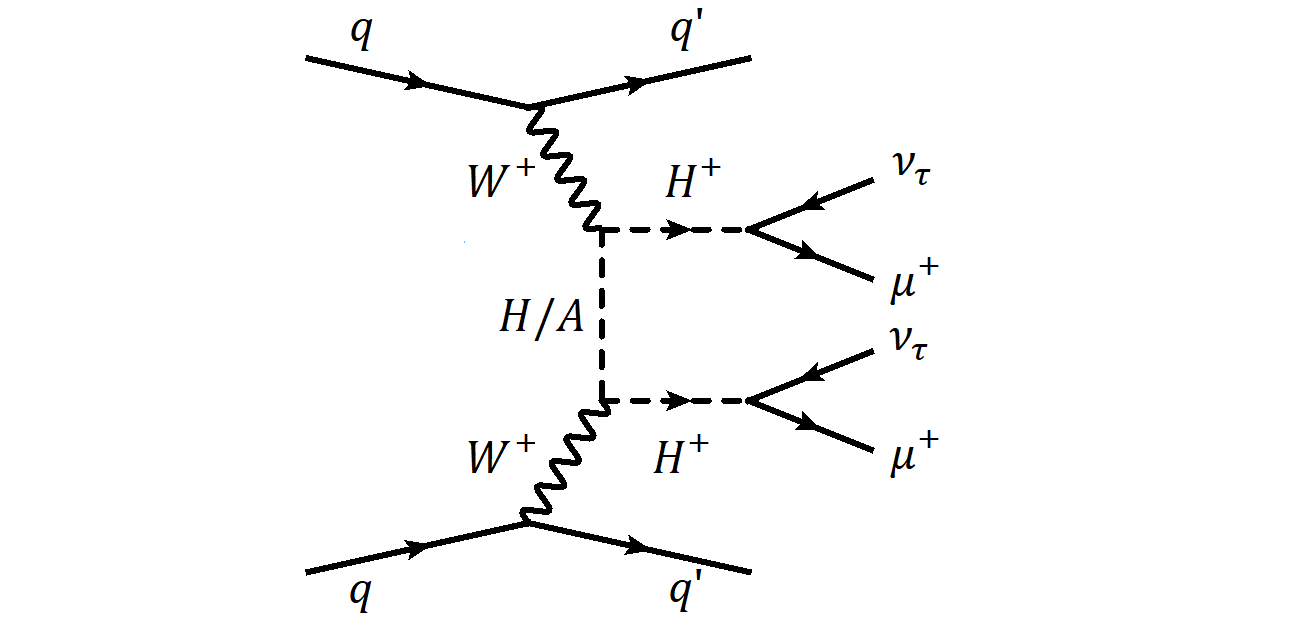} 
$$
\caption{Feynman diagram for the signal $pp \rightarrow \mu^+\mu^+ jj +$ ${E\!\!\!\!/}_{T}$  at the LHC.} \label{lnv}
\end{figure}

\begin{figure}[htb!]
$$
\includegraphics[width=0.46\textwidth]{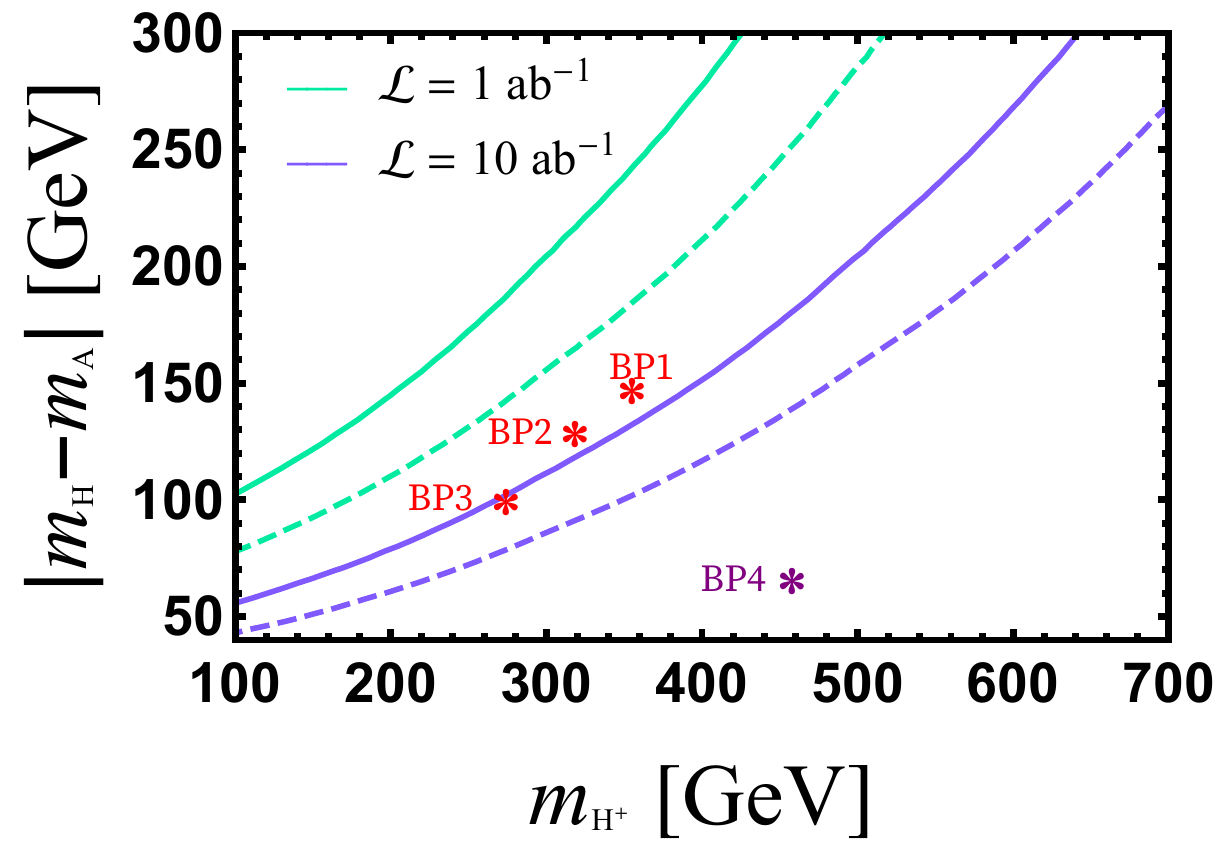}
$$
\caption{Discovery prospect of the signal $pp \rightarrow \mu^+\mu^+ jj +$ ${E\!\!\!\!/}_{T}$  at the 14 TeV LHC  in the mass-splitting vs. charged Higgs mass plane. Dashed and solid lines correspond to a significance of 3$\sigma$ and 5 $\sigma$.} \label{reach}
\end{figure}

\begin{table*}[th]
\centering
\begin{ruledtabular}
\begin{tabular}{ccccccc}
$\sqrt{| \tilde{Y}_{\mu\tau}  \tilde{Y}_{\tau\mu} |}$&  & $m_H$ (GeV) &
$m_A$ (GeV) &
$m_{H^{\pm}}$ (GeV) &
$M_{W}$ (GeV) & $\Delta a_{\mu}\times 10^{9}$   \\
\hline 
& BP1 & 445.15  & 343.42 & 274.21 & 80.4372 &  3.04
\\\cline{2-7}
0.3& BP2 &372.48  & 503.42 & 318.88 & 80.4241 &  2.95
\\\cline{2-7}
& BP3 & 402.50  & 551.76 & 355.12 & 80.4203 &  2.66
\\\hline
& BP4 & 593.90  & 526.20 & 457.68 & 80.4227 &  $2.08$
\\\cline{2-7}
0.5& BP5 &397.48  & 430.29 & 511.94 & 80.4293 &  2.32
\\\cline{2-7}
& BP6 & 128.68  & 126.95 & 233.72 & 80.4476 &  $2.98$
\\ \hline
&BP7 & 508.26  & 494.50 & 598.65 & 80.4292 & $2.28$
\\\cline{2-7}
1.0 & BP8 &400.32  & 409.43 & 501.73 & 80.4294 & $2.73$
\\\cline{2-7}
& BP9 & 575.44  & 555.96 & 466.60 & 80.4257 & 2.31
\\
\end{tabular}
\end{ruledtabular}
\caption{Benchmark points that simultaneously fit  the $W$-boson mass ($M_W$) and muon anomalous magnetic moment ($\Delta a_{\mu}$) within the model, along with the predicted values of $M_W$ and $\Delta a_\mu$.
\label{benchmarkpoints}
}
\end{table*}
{\textbf {\textit {Collider tests.--}}} Since the masses of the neutral and charged scalar cannot be more than about 600 GeV in order to explain the $W$-boson mass and the muon $(g-2)$ anomaly, the model can be tested at the LHC with the high luminosity run. One interesting process is $pp \rightarrow (\mu^+ \mu^+jj+ \slashed{E}_T)$ which arises from the vector boson fusion (VBF) diagram shown in Fig. \ref{lnv} \cite{Aiko:2019mww}. The amplitude for this process is proportional to the mass-splittings among the neutral scalar and pseudoscalar particles, which in turn is constrained by muon $(g-2)$ and $W$-boson mass shift. Now, we analyze the signal  $pp \rightarrow \mu^+\mu^+ jj +$ ${E\!\!\!\!/}_{T}$ (see Appendix, for analysis details) and summarize the signal sensitivity in  Fig.\ref{reach}  at 14 $\mathrm{TeV}$ LHC for two different integrated luminosities. For $|m_H - m_A| = 200\, (150)$ GeV, we find  that at 3$\sigma$ level, the charged scalar $H^+$ can be probed up to masses of  382 (289), and 591  (485) GeV respectively for the integrated luminosities of $\mathcal{L}=1$ and 10 ab$^{-1}$. It is worth noting that mass splitting is necessary to account for the proper sign and strength of the $W-$ mass shift, although, as shown by the benchmark points, $\mathcal{O} (10)$ GeV mass splitting is adequate to account for the shift. However, the sensitivity of the signal $pp \rightarrow \mu^+\mu^+ jj +$ ${E\!\!\!\!/}_{T}$ at the 14 TeV LHC is insufficient to test the scenario in the case of such small mass-splittings. However, in the complementary parameter space, when mass splitting is significant ($\gtrsim 50$ GeV) between $H$ and $A$, as seen in Fig.\ref{reach}, this may be investigated during the forthcoming run of the LHC. For instance, BP1, BP2, and BP3 can be fully explored with future HL-LHC with 10 ab$^{-1}$ luminosity.


LHC could also look for pair production of $H$ and $A$ via $Z$-boson exchange. Each of these fields would decay into $\mu^\pm \tau^\mp$.  No dedicated search has been carried out by the LHC collaborations.  The signal strength for this process has been studied in detail in Ref. \cite{Babu:2020ivd} and Ref. \cite{Iguro:2019sly}, where it has been shown that current data already would be sensitive to $H$ and $A$ masses of about 500 GeV. For completeness, we analyze the signal using the same technique as described in Ref.~\cite{Babu:2020ivd}. After imposing the acceptance criteria specified in  Ref.~\cite{Babu:2020ivd}, we estimate that the number of events for the signal that passes the acceptance criteria is 185, 270, 440, 90, 360, 39100, 148 ,390, 80 for the nine benchmark points, respectively, whereas  we obtain 560 events for the SM background at 14 TeV LHC with an integrated luminosity of $\mathcal{L}=$ 1 ab$^{-1}$. We find that all of these benchmark points can be explored with a significance of more than  $3 \sigma $  at 14 TeV LHC with an integrated luminosity of $\mathcal{L}=$ 1 ab$^{-1}$. For instance, the significance values associated with the nine benchmark points are 6.7, 9.3, 13.9, 3.5, 11.8, 196, 5.5, 12.6, and 3.2, respectively. Thus the scenario presented can be fully explored in future searches at the LHC.

{\textbf {\textit {Conclusions.--}}} In this paper we have shown an intriguing connection between the recently reported upward shift in the $W$-boson mass measurement by the CDF collaboration and the muon $(g-2)$ anomaly within the 2HDM.  Mass splittings among the second Higgs multiplet are needed in the scenerio presented in order to explain the muon $(g-2)$.  These splittings also result in an upward shift of $M_W$, consistent with the CDF results.  The charged and neutral scalars of the second Higgs doublet should have masses below about 600 GeV, which makes the scenario testable at the high luminosity LHC.  We have pointed out that the mass splittings among the scalars also would lead to the signal  $pp \rightarrow (\mu^+ \mu^+jj+ \slashed{E}_T)$ which arises from the vector boson fusion diagrams. Additionally, $pp \rightarrow (\mu^+\mu^-\tau^+\tau^-+X)$ can be looked for, with signal strength within reach for the entire range of scalar mass parameters.

\vspace{0.1in}
\begin{acknowledgments}
{\textbf {\textit {Acknowledgments.--}}} The work of KSB and VPK is in part supported by US Department of Energy Grant Number DE-SC 0016013. SJ acknowledges Oklahoma State Univeristy physics department for hospitality while this work was done.

 \end{acknowledgments}
 
 \begin{widetext}
		\appendix
		\section{Scalar potential}
\noindent The most general scalar potential in the Higgs basis can be written as
\begin{align}
&V= m_{11}^2H_1^{\dagger}H_1+m_{22}^2H_2^{\dagger}H_2
-\{m_{12}^2H_1^{\dagger}H_2+{\rm h.c.}\} 
+\frac{\lambda_1}{2}(H_1^{\dagger}H_1)^2
+\frac{\lambda_2}{2}(H_2^{\dagger}H_2)^2
+\lambda_3(H_1^{\dagger}H_1)(H_2^{\dagger}H_2)
\nonumber\\ &
+\lambda_4(H_1^{\dagger}H_2)(H_2^{\dagger}H_1)
+\left\{\frac{\lambda_5}{2}(H_1^{\dagger}H_2)^2+{\rm h.c.}\right\}
+\left\{
\big[\lambda_6(H_1^{\dagger}H_1)
+\lambda_7(H_2^{\dagger}H_2)\big]
H_1^{\dagger}H_2+{\rm h.c.}\right\}.
\label{pot}
\end{align}
Minimization of the potential leads to the conditions $m_{11}^2+\frac{\lambda_1}{2}v^2=0$ and $m_{12}^2-\frac{\lambda_6}{2}v^2 = 0$. 

		\section{Expression for $S$-parameter}
In the alignment limit, the $S$-parameter in the 2HDM is given by \cite{Barbieri:2006dq}
 \begin{align} 
    & S  =  \dfrac{1}{2 \pi}\int_0^1 dx\, x(1-x)\ln{\left(\frac{x m_H^2+(1-x)m_A^2}{m^2_{H^{\pm}}}\right)} . \label{eq:S}
\end{align}

		\section{Correction to $G_F$ in muon decay}
	The  shift in the $G_F$ affecting muon decay due to one-loop wave function  corrections to the $\mu$ and $\nu_\mu$ lines is given by \cite{Abe:2017jqo}
 \begin{align} 
    & \frac{\delta G_F}{G_F}  \simeq \frac{|\tilde{Y}_{\mu\tau}|^2}{32\pi^2}\left(1-f(m_H^2,m^2_{H^{\pm}})-f(m_A^2,m^2_{H^{\pm}})\right)\, ,  \label{eq:GF}
\end{align}
where the function $f$ is defined as 
\begin{equation} \label{Fdef}
    f(m_1^2,m_2^2) \  \equiv \ \frac{m_1^2+m^2_{2}}{4(m_1^2-m^2_{2})}\ln{\frac{m_1^2}{m^2_{2}}} \,.
\end{equation}

	\section{Details on collider analysis}
\noindent For our collider analysis, we analyze the signal  with {\sc MadGraph5aMC@NLO}~\cite{Alwall:2011uj,Alwall:2014hca} event generator, simulating the hadronization and underlying event effects with {\sc Pythia8}~\cite{Sjostrand:2007gs} and detector effects with the {\sc Delphes3}~\cite{deFavereau:2013fsa}.  For the signal $pp \rightarrow (\mu^+ \mu^+jj+ \slashed{E}_T)$, the dominant SM background originates from same-sign $W-$boson pair-production. In order to optimize the signal sensitivity, we employ the following selection criterias: (a) events with at least two jets and two same-sign dimuon pairs with $p_{Tj}>30$~GeV, $|\eta^j|<5.0$, $p_{T}^{\mu}>~20 \mathrm{GeV}$ and $\left|\eta^{\mu}\right|<2.5$ are selected; (b) for the two leading jets, invariant mass cut $m_{j j}>500~\mathrm{GeV}$ and pseudorapidity separation cuts $\left|\Delta \eta_{j j}\right|>2.5$ are imposed.  We define the significance as $S/\sqrt{S+B}$, where $S$ and $B$ define the number of events for signal and SM backgrounds. 
\end{widetext}

\vspace{-0.398in}
\bibliographystyle{utphys}
\bibliography{reference}

\end{document}